\documentclass[useAMS,usenatbib]{mn2e}              
\usepackage{epsfig}
\newcommand{\kms}{\>{\rm km}\,{\rm s}^{-1}}

\newcommand{\beq}{\begin{equation}}
\newcommand{\eeq}{\end{equation}}

\newcommand{\apj}{ApJ}
\newcommand{\apjl}{ApJL}
\newcommand{\apjs}{ApJS}
\newcommand{\aj}{AJ}
\newcommand{\mnras}{MNRAS}
\newcommand{\aap}{A\&A}

\newcommand{\araa}{ARA\&A}
\newcommand{\pasp}{PASP}

\newcommand{\physrep}{Physics Report}

\newdimen\hssize
\newdimen\hdsize 
\title[Clustering of star-forming  galaxies] 
{Interaction-induced star formation in a complete sample of 
$10^5$ nearby star-forming galaxies}  
\author[Li et al.]
{Cheng Li$^{1,2}$\thanks{E-mail: leech@mpa-garching.mpg.de},
Guinevere Kauffmann$^{2}$,  
Timothy M. Heckman$^{3}$,
Y. P. Jing$^{1}$, 
\newauthor Simon D. M. White$^{2}$
\\  
${^1}$ MPA/SHAO  Joint  Center  for  Astrophysical Cosmology 
       at Shanghai Astronomical Observatory, 
       Nandan Road 80, Shanghai 200030, China \\ 
${^2}$ Max Planck Institut f\"ur Astrophysik, 
       Karl-Schwarzschild-Strasse 1, 85748 Garching, Germany \\
${^3}$ Department of Physics and Astronomy,
       Johns Hopkins University, Baltimore, MD 21218
}

\begin{document}

\graphicspath{{figs/}}

\date{Accepted ........ Received ........; in original form ........}

\pagerange{\pageref{firstpage}--\pageref{lastpage}} \pubyear{2007}

\maketitle

\label{firstpage}

\begin  {abstract}  We  investigate  the clustering  properties  of  a
complete sample  of $10^5$ star-forming  galaxies drawn from  the data
release 4 (DR4) of the Sloan  Digital Sky Survey.  On scales less than
100 kpc , the amplitude  of the correlation function exhibits a strong
dependence  on the  specific star  formation rate  of the  galaxy.  We
interpret this as the signature  of enhanced star formation induced by
tidal interactions.   We then explore  how the average  star formation
rate in a galaxy is enhanced as the projected separation $r_p$ between
the galaxy and its companions decreases.  We find that the enhancement
depends strongly on $r_p$, but  very weakly on the relative luminosity
of  the companions.   The enhancement  is  also stronger  in low  mass
galaxies than  in high mass galaxies.   In order to  explore whether a
tidal  interaction  is not  only  sufficient,  but  also necessary  to
trigger  enhanced star formation  in a  galaxy, we  compute background
subtracted  neighbour counts  for  the galaxies  in  our sample.   The
average number of close neighbours around galaxies with low to average
values of  SFR/$M_*$ is close to  zero.  At the  highest specific star
formation rates, however, more than 40\% of the galaxies in our sample
have  a  companion within  a  projected  radius  of 100  kpc.   Visual
inspection  of  the  highest  SFR/$M_*$  galaxies  without  companions
reveals that more than 50\%  of these are clear interacting or merging
systems.  We conclude that tidal interactions are the dominant trigger
of enhanced star formation  in the most strongly star-forming systems.
Finally, we find clear evidence  that tidal interactions not only lead
to  enhanced star  formation in  galaxies, but  also  cause structural
changes such as an increase in concentration.

\end {abstract}


\section {Introduction}

It has been known for  more than thirty years that galaxy interactions
lead  to enhanced  star formation.   \cite{Toomre-Toomre-72} pioneered
the use of numerical simulations to study the interactions of galaxies
and suggested that gas may be  funnelled to the central regions of the
systems as a result of the strong tidal forces that operate during the
encounter. This  gas is then able  to fuel a burst  of star formation.
Since  then, there  have  been many  studies,  both observational  and
theoretical,  that   have  examined  the   relationship  between  star
formation and galaxy interactions.

Most   early  observational   studies  adopted   broad   band  colours
\citep[e.g.][]{Larson-Tinsley-78},    H$\alpha$   equivalent   widths,
\citep[e.g.][]{Keel-85,Bushouse-86,Kennicutt-87},          far-infrared
luminosities   \citep[e.g.][]{Bushouse-Werner-Lamb-88},  or  molecular
(CO) emission \citep{Young-86, Sanders-86, Solomon-Sage-88, Tinney-90,
Young-96}  as  indicators  of   star  formation.   These  studies  all
demonstrated  that  galaxy interactions  are  statistically linked  to
enhanced   rates    of   star   formation    \citep[see   the   review
of][]{Keel-91,Struck-99}.

Recent  studies of star  formation in  interacting galaxies  have been
based on redshift surveys such as the Center for Astrophysics redshift
survey
\citep[CfA2;][]{Barton-Geller-Kenyon-00,Woods-Geller-Barton-06},    the
Two Degree Field Redshift Survey \citep[2dFGRS;][]{Lambas-03}, and the
Sloan                 Digital                Sky                Survey
\citep[SDSS;][]{Nikolic-Cullen-Alexander-04,Woods-Geller-07,Ellison-08}.
These  studies have  also  provided observational  evidence that  star
formation is enhanced as a consequence of tidal interactions.  Most of
these studies have also demonstrated that the degree of enhancement is
a strong function of the projected separation between the two galaxies
as well  as their difference  in redshift.  In addition,  some studies
investigated   how    galaxy   properties   such    as   concentration
\citep{Nikolic-Cullen-Alexander-04},          luminosity         ratio
\citep{Woods-Geller-Barton-06,Woods-Geller-07},   stellar  mass  ratio
\citep{Ellison-08},  colour, and AGN  activity \citep{Woods-Geller-07}
depend on separation.

Although  most studies  have supported  the picture  that interactions
induce  star formation, there  have been  number of  dissenting papers.
For     example,     \citet{Bergvall-Laurikainen-Aalto-03}    analyzed
optical/near-IR  observations of  a sample  of  59 interacting/merging
systems and concluded that they  do not differ very much from isolated
galaxies   in   terms   of   their  global   star   formation   rates.
\citet{Brosch-Almoznino-Heller-04} found that interaction-induced star
formation is not significant for  dwarf galaxies.  A more recent study
by \citet{Smith-07}  analyzed Spitzer mid-infrared (MIR)  imaging of a
sample  of 35  interacting galaxy  pairs selected  from the  Arp Atlas
\citep{Arp-66}.   They compared  the  global MIR  properties of  these
systems  with those  of normal  spiral  galaxies.  The  MIR colors  of
interacting  galaxies were  found to  be redder  than  normal spirals,
implying enhancements  to the  specific SFRs of  a factor  of $\sim$2.
However, in contrast to results from previous investigations, they did
not   find   any   evidence   that   the   enhancement   depended   on
separation. This may be due to the small size of their sample and fact
that   the   galaxies   were   selected  to   be   tidally   disturbed
\citep{Smith-07}.

On  the   theoretical  side,  $N$-body  simulations   that  treat  the
hydrodynamics      of     the      gas     \citep{Negroponte-White-83,
Barnes-Hernquist-92,   Mihos-Hernquist-96,   Springel-00,  Tissera-02,
Meza-03,  Kapferer-05,  Cox-06}  have demonstrated  that  interactions
between galaxies can bring gas from the disc to the central regions of
the  galaxy,  leading  to   enhanced  star  formation  in  the  bulge.
Recently, \citet{DiMatteo-07}  investigated star formation  in a suite
of several hundred numerical  simulations of interacting galaxies with
different gas fractions,  bulge-to-disk ratios and orbital parameters.
Their work confirmed that  galaxy interactions and mergers can trigger
strong nuclear starbursts.  However, the authors pointed out that this
is not always the case, because strong tidal interactions at the first
pericenter passage  can remove a large  amount of gas  from the galaxy
disks, and this  gas is only partially re-acquired  by the galaxies in
the last phase of the merging event.

In   summary,    although   it   is   now    well   established   that
interactions/mergers   between  galaxies   {\em   can}  enhance   star
formation, a number of important questions remain to be answered:
\begin{itemize}
\item  Are interactions  not  only sufficient  but  also necessary  to
enhance star formation?
\item Do interactions {\em always} trigger enhanced star formation?
\item How does the enhancement  in star formation depend on parameters
such as the separation between  the two galaxies and their mass ratio?
Does the enhancement also depend on properties such as stellar mass or
galaxy morphology?
\end{itemize}

To answer these questions, we adopt three different methods to analyse
a sample  of $\sim10^5$ star-forming  galaxies selected from  the Data
Release 4 (DR4) of the  SDSS.  First, we compute the cross-correlation
between star-forming galaxies and a reference sample of galaxies drawn
from  the DR4.   In the  standard  model of  structure formation,  the
amplitude of the correlation function  on scales larger than a few Mpc
provides a  direct measure of the  mass of the dark  matter haloes that
host the galaxies.  As we  will show, the amplitude of the correlation
function on  scales less than $\sim 100$  kpc can serve as  a probe of
physical processes such as  mergers and interactions.  We then compute
the  average  enhancement in  star  formation  as  a function  of  the
projected  separation between  two  galaxies and  we  explore how  the
enhancement  depends on  galaxy properties  such as  stellar  mass and
concentration index.   Finally, we compute counts  around our galaxies
as a function of separation and explore how this changes as a function
of  the specific  star formation  rate  SFR/$M_*$. This  allows us  to
investigate  whether  the  majority  of galaxies  with  specific  star
formation  rates  above   some  critical  threshold  are  experiencing
merger-induced starbursts.   In a  separate paper, we  explore whether
AGN activity  is also triggered  by tidal interactions using  the same
set of analysis techniques.

Throughout this paper, We assume a cosmological model with the density
parameter $\Omega_0=0.3$ and  a cosmological constant $\Lambda_0=0.7$.
To avoid the $-5\log_{10}h$ factor,  a Hubble constant $h=1$, in units
of  $100\kms{\rm Mpc}^{-1}$,  is  assumed throughout  this paper  when
computing absolute magnitudes.

\section{Samples}

\subsection {The SDSS Spectroscopic Sample}

The data analyzed  in this study are drawn from  the Sloan Digital Sky
Survey (SDSS).  The survey goals are to obtain photometry of a quarter
of  the sky and  spectra of  nearly one  million objects.   Imaging is
obtained    in     the    {\em    u,    g,    r,     i,    z}    bands
\citep{Fukugita-96,Smith-02,Ivezic-04}  with a  special  purpose drift
scan camera  \citep{Gunn-98} mounted  on the SDSS  2.5~meter telescope
\citep{Gunn-06}  at Apache  Point Observatory.   The imaging  data are
photometrically    \citep{Hogg-01,Tucker-06}    and    astrometrically
\citep{Pier-03} calibrated,  and used  to select stars,  galaxies, and
quasars  for follow-up fibre  spectroscopy.  Spectroscopic  fibres are
assigned to  objects on  the sky using  an efficient  tiling algorithm
designed to optimize  completeness \citep{Blanton-03}.  The details of
the survey strategy can be found in \citet{York-00} and an overview of
the data pipelines and products  is provided in the Early Data Release
paper  \citep{Stoughton-02}. More details  on the  photometric pipeline
can be found in \citet{Lupton-01}.

Our parent sample for this  study is composed of 397,344 objects which
have  been  spectroscopically  confirmed  as galaxies  and  have  data
publicly     available     in      the     SDSS     Data     Release~4
\citep{Adelman-McCarthy-06}.   These  galaxies are  part  of the  SDSS
`main'  galaxy   sample  used   for  large  scale   structure  studies
\citep{Strauss-02}  and have  Petrosian  $r$ magnitudes  in the  range
$14.5 < r < 17.77$ after correction for foreground galactic extinction
using  the  reddening  maps  of  \citet{Schlegel-Finkbeiner-Davis-98}.
Their redshift  distribution extends from $\sim0.005$ to  0.30, with a
median $z$ of 0.10.

The SDSS spectra are obtained with two 320-fibre spectrographs mounted
on  the SDSS  2.5-meter telescope.   Fibers 3  arcsec in  diameter are
manually plugged  into custom-drilled  aluminum plates mounted  at the
focal plane of  the telescope. The spectra are  exposed for 45 minutes
or until a fiducial signal-to-noise  (S/N) is reached.  The median S/N
per pixel  for galaxies in the  main sample is  $\sim14$.  The spectra
are  processed by  an automated  pipeline, which  flux  and wavelength
calibrates  the   data  from  3800  to   9200~\AA.   The  instrumental
resolution  is   R~$\equiv  \lambda/\delta\lambda$  =   1850  --  2200
(FWHM$\sim2.4$~\AA\ at 5000~\AA).

\subsection{Star-forming galaxies}

Our   sample  of  star-forming   galaxies  is   drawn  from   the  DR4
spectroscopic    sample    using    the    criteria    described    in
\citet{Brinchmann-04}. In order for a galaxy to be securely classified
as  star-forming, we  require  that the  four  emission lines  [OIII],
H$\beta$,  H$\alpha$ and  [NII] all  be detected  with signal-to-noise
greater than 3 and that the ratios [OIII]/H$\beta$ and [NII]/H$\alpha$
have   values   that   place   them   within   the   region   of   the
\citet[][BPT]{Baldwin-Phillips-Terlevich-81}   diagram   occupied   by
galaxies in which  the primary source of ionizing  photons is from HII
regions rather  than an AGN. We refer  to this sample as  the high S/N
star-forming class.   In certain cases, we supplement  the sample with
the low S/N star-forming class defined by Brinchmann et al.  These are
the  galaxies that  are  left over  after  all the  AGN  and high  S/N
star-forming galaxies  have been  removed, and they  have S/N  $>2$ in
H$\alpha$.   Star formation rates  can still  be estimated  from their
emission line  strengths, but  the errors on  these estimates  will be
significantly larger than for the high S/N sample.

The  reader  is  referred  to  \citet{Brinchmann-04}  for  a  detailed
description of  how star formation  rates are derived for  the various
samples. We  will be  making use of  the specific star  formation rate
$SFR/M_*$ estimated within the  3 arsecond SDSS fibre aperture.  These
star formation rates  are more accurate than the  total star formation
rates derived  by Brinchmann  et al, because  they depend only  on the
emission line fluxes measured from the spectra and they do not involve
any uncertain colour corrections.  The disadvantage of the fibre-based
specific star formation  rates is that they are  only sensitive to the
emission from the inner region of the galaxy, which includes one third
of the total light on average.

\subsection{Reference Samples}

We   work  with   two  different   reference  samples:   (i)   a  {\em
spectroscopic}  reference  sample,  which   is  used  to  compute  the
projected  cross-correlation function $w_p(r_p)$  between star-forming
galaxies  and  reference  galaxies,   and  (ii)  a  {\em  photometric}
reference  sample,  which  is   used  to  calculate  counts  of  close
neighbours  around  star-forming  galaxies.    We  use  the  New  York
University Value  Added Galaxy  Catalogue (NYU-VAGC) to  construct the
reference  samples. The  original  NYU-VAGC is  a  catalogue of  local
galaxies    (mostly    below    $z\approx   0.3$)    constructed    by
\cite{Blanton-05} based on the SDSS DR2. Here, we use a new version of
the  NYU-VAGC ({\tt  Sample dr4}),  which is  based on  SDSS  DR4. The
NYU-VAGC is described in detail in \cite{Blanton-05}.

The reference samples  are exactly the same as  used in \cite{Li-06b}.
In  short,  the  spectroscopic  reference  sample  is  constructed  by
selecting from {\tt  Sample dr4} all galaxies with $14.5 <  r < 17.6 $
that are identified as galaxies  from the Main sample, in the redshift
range    $0.01\leq   z\leq0.3$,    and   with    absolute   magnitudes
$-23<M_{^{0.1}r}<-17$.   The spectroscopic  reference  sample contains
292,782   galaxies.   The   photometric  reference   sample   is  also
constructed  from {\tt  Sample  dr4} by  selecting  all galaxies  with
$14.5<r<19$.   The resulting sample  includes 1,065,183  galaxies.  In
certain cases, we will work  with photometric reference samples with a
range of differing limiting magnitudes.

\section{Cross-correlation functions}

\begin{figure} 
\centerline{\psfig{figure=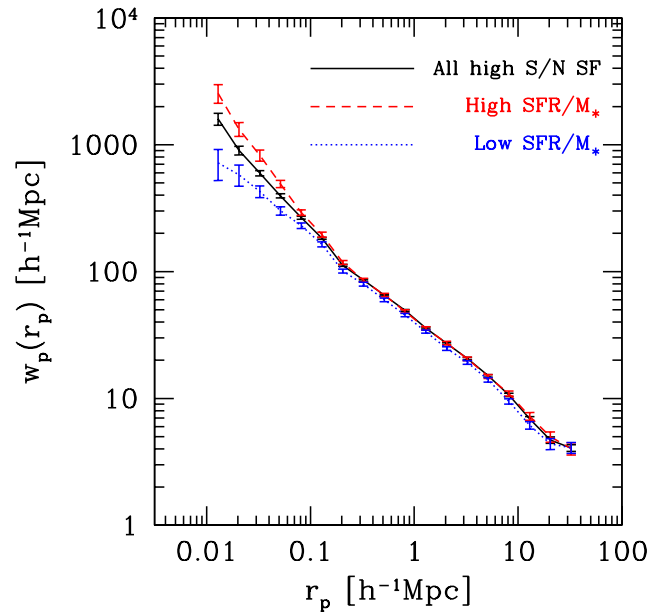,clip=true,width=0.5\textwidth}}
\caption{Projected  redshift-space 2-point  cross-correlation function
$w_p(r_p)$  between  star-forming galaxies  and  the reference  galaxy
sample.   Different  lines correspond  to  star-forming galaxies  with
different specific star formation rates.   See the text for a detailed
description.}
\label{fig:wrp}
\end{figure}

\begin{figure*}      
\centerline{\psfig{figure=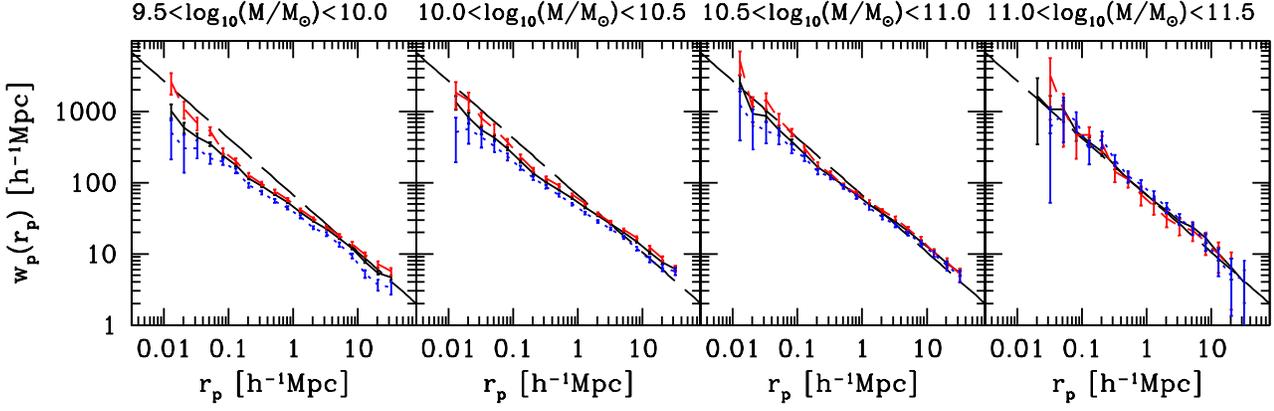,clip=true,width=\textwidth}}
\caption{Similar to  Figure~1, but  in different intervals  of stellar
mass as indicated at the top  of the figure.  The symbols are the same
as in  Figure~\ref{fig:wrp}, except that a power  law corresponding to
$\xi(r)=(r/5h^{-1}Mpc)^{-1.8}$ is  additionally plotted in  each panel
as a long-dashed line.}
\label{fig:wrp_smass}
\end{figure*}

\begin{figure*}
\centerline{\psfig{figure=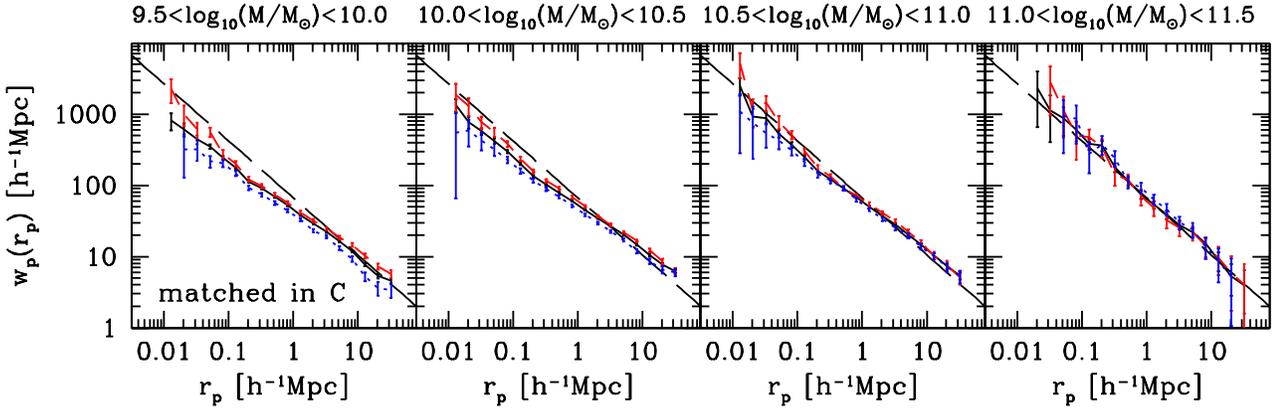,clip=true,width=\textwidth}}
\caption{Same  as Figure~\ref{fig:wrp_smass},  except  that the  three
$SFR/M_\ast$ samples in each panel are matched in concentration.}
\label{fig:wrp_smass_c}
\end{figure*}

\begin{figure*}
\centerline{\psfig{figure=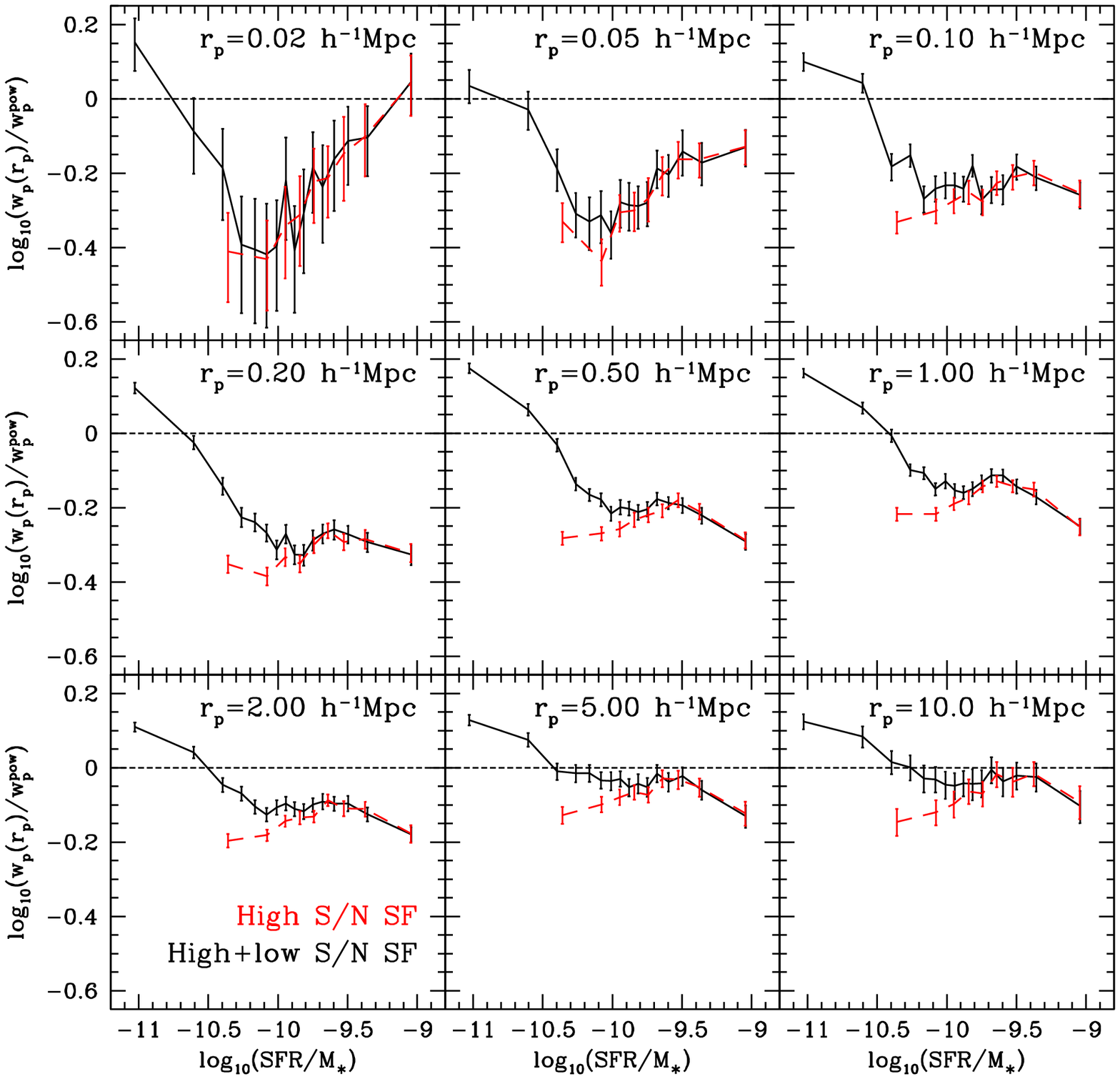,clip=true,width=\textwidth}}
\caption{The projected  2PCCF $w_p(r_p)$  normalized by the  power law
corresponding to a  real-space 2PCF of $\xi(r)=(r/5h^{-1}Mpc)^{-1.8}$,
as  measured  at  different  physical  scales and  as  a  function  of
$SFR/M_\ast$.   Dashed lines  are for  high S/N  star-forming galaxies
only, while  solid lines  show results for  the sample  including both
high and low S/N star-forming galaxies.}
\label{fig:wrp_sfr}
\end{figure*}

Our methodology for computing correlation functions has been described
in detail  in our  previous papers \citep{Li-06a,Li-06b}.   We present
here a  brief description  and the reader  is referred to  the earlier
papers for details.  Random samples are constructed that have the same
selection  function  as  the  reference  sample.   The  redshift-space
two-point  cross-correlation function  (2PCCF)  $\xi(r_p,\pi)$ between
star-forming  galaxies and  the  reference sample  is then  calculated
using  the  estimator   presented  in  \citet{Li-06b}.   Finally,  the
redshift-space projected 2PCCF  $w_p(r_p)$ is estimated by integrating
$\xi(r_p,\pi)$  along the line-of-sight  direction $\pi$  with $|\pi|$
ranging from  0 to 40  $h^{-1}Mpc$.  We have also  corrected carefully
for the effect of fibre collisions  and a description and tests of the
method  are given  in \citet{Li-06b}.   The errors  on  the clustering
measurements  are estimated using  the bootstrap  resampling technique
\citep{Barrow-Bhavsar-Sonoda-84}.

We first  compute $w_p(r_p)$ for  our sample of high  S/N star-forming
galaxies from the SDSS DR4. In order to study how this depends on star
formation rate (SFR),  we rank all the high  S/N star-forming galaxies
according to the values of their specific star formation rates (SSFR),
$SFR/M_\ast$,  and define  subsamples of  'high SSFR'  and  'low SSFR'
galaxies  as   those  contained  within  the  upper   and  lower  25th
percentiles  of the distribution  of this  quantity.  The  results are
shown in  Figure~\ref{fig:wrp}.  The dashed  (dotted) line corresponds
to  the high  (low) SSFR  subsample, while  the solid  line  shows the
result for the sample as a whole.

Figure~\ref{fig:wrp} shows that galaxies with higher $SFR/M_\ast$ have
stronger  clustering on  scales smaller  than 0.1  Mpc and  the effect
becomes stronger at smaller  projected separations.  As pointed out by
\citet{Li-06b},  the clustering  amplitude  of galaxies  depends on  a
variety  of  galaxy  properties,  including stellar  mass  and  galaxy
structure.  If we  wish to  isolate the  effect of  the  specific star
formation  rate, it is  important that  we make  sure that  the galaxy
samples  that  we  study  are   closely  matched  in  terms  of  other
properties,  so that  the effect  on the  star formation  rate  can be
isolated. We have thus divided  all the high S/N star-forming galaxies
into  four subsamples  according to  $\log_{10}(M_\ast/M_\odot)$.  For
each  subsample  we repeat  the  clustering  analysis  as above.   The
results  are  shown in  Figure~\ref{fig:wrp_smass}.   The four  panels
correspond to different  intervals of $\log_{10}(M_\ast/M_\odot)$.  To
guide  the eye,  a power  law corresponding  to a  real-space  2PCF of
$\xi(r)=(r/5h^{-1}Mpc)^{-1.8}$  is plotted  as a  long-dashed  line in
each panel.   We see  that the amplitude  of $w_p(r_p)$  increases for
galaxies  with larger  stellar masses.   This is  consistent  with our
previous findings about the  mass dependence of galaxy clustering.  We
also see that the difference  in clustering between galaxies with high
and low SFR/$M_*$ on scales smaller than 0.1 Mpc is most pronounced in
the lowest stellar mass interval.   Next, in each of the four $M_\ast$
intervals, we  match the  three $SFR/M_\ast$ samples  in concentration
parameter $C$ by requiring that the distribution of $C$ is exactly the
same as in each of these samples. The $w_p(r_p)$ measurements for such
matched  samples  are   shown  in  Figure~\ref{fig:wrp_smass_c}.   The
results are very similar to those shown in the previous figure.

We  conclude that  the  small scale  clustering  dependences shown  in
Figure~\ref{fig:wrp} are  genuinely related to  the differing specific
star   formation   rates   of    the   galaxies   in   the   different
samples.  Galaxies  with the  highest  specific  star formation  rates
apparently have  an excess of companions  on scales less  than 100 kpc
when compared  to the average  star-forming galaxy. The fact  that the
increase in clustering occurs only  on very small scales suggests that
the  excess   star  formation  is   being  triggered  by   {\em  tidal
interactions} with these  companions.  Another intriguing result shown
in these figures is that  galaxies with low $SFR/M_\ast$ are {\em less
clustered} on small  scales. This suggests that there  might be a {\em
continuous}  trend  linking average  number  of  close neighbours  and
$SFR/M_\ast$.

To investigate  this in more  detail, we calculate how  the clustering
amplitude depends  on $SFR/M_\ast$ at a variety  of different physical
scales.  The results are shown in Figure~\ref{fig:wrp_sfr} (red dashed
lines).  One problem with the  high S/N star-forming sample is that it
does not  extend to $SFR/M_\ast$  values much below $\sim  -10.5$.  To
extend our analysis to lower values, we include the sample of {\em low
S/N}  star-forming  galaxies  defined  by  \citet{Brinchmann-04}.   As
discussed in section  2.2, the star formation rates  in these galaxies
are  estimated  from  the  H$\alpha$  line luminosity,  but  the  dust
correction is quite uncertain because H$\beta$ is not usually detected
with high S/N.   Results where the low S/N  star-forming galaxies have
been    included   are    plotted    as   black    solid   lines    in
Figure~\ref{fig:wrp_sfr}.  As can be seen, $\log SFR/M_*$ extends down
to values around $\sim -11$ for this sample.

On  scales larger than  100 kpc,  there is  very little  dependence of
clustering amplitude on specific star formation rate for $\log SFR/M_*
>  -10$.   At lower  values  of  $SFR/M_*$,  the clustering  amplitude
increases.   This is a  manifestation of  the strong  relation between
star  formation and local  density or  environment.  It  is well-known
that galaxies  located in dense,  massive structures such  as clusters
have  lower  specific star  formation  rates  than ``field''  galaxies
\citep[e.g.][]{Kauffmann-04}. It  is currently accepted  that after a
galaxy  is  accreted onto  a  larger structure,  such  as  a group  or
cluster, its star formation rate  will decline, either because its gas
is  removed by  processes such  as ram-pressure  stripping,  or simply
because no further  gas accretion takes place and  the galaxy runs out
of the fuel to make new stars.

On scales less than 100 kpc, the dependence of clustering amplitude on
specific star formation  rate is more complicated. At  values of $\log
SFR/M_*$  less  than -10,  we  see  the  same increase  in  clustering
amplitude  that we  saw on  larger  scales. This  may appear  somewhat
surprising at first. In a recent paper, however, \citet{Barton-07} use
cosmological  simulations  to  show  that a  substantial  fraction  of
galaxies selected as ``close pairs'' from surveys such as SDSS or 2DF,
do in  fact reside in very  massive dark matter halos.   Based on this
work, we conjecture that the rise in clustering amplitude seen at all
separations  at low values  of SFR/$M_*$  rate is  the result  of star
formation shutting down  in galaxy groups and clusters.   At values of
$\log SFR/M_*$ greater than -10 and at separations less than $\sim 50$
kpc,  the clustering  amplitude  shows a  strong  and continuous  {\em
increase} towards larger values of  $SFR/M_*$.  This is a clear signal
that mergers or interactions play an important role in triggering {\em
enhanced} star formation in galaxies.

\section{Star formation enhancement functions}
\label{sec:enhancement}

\begin{figure*}
\centerline{\psfig{figure=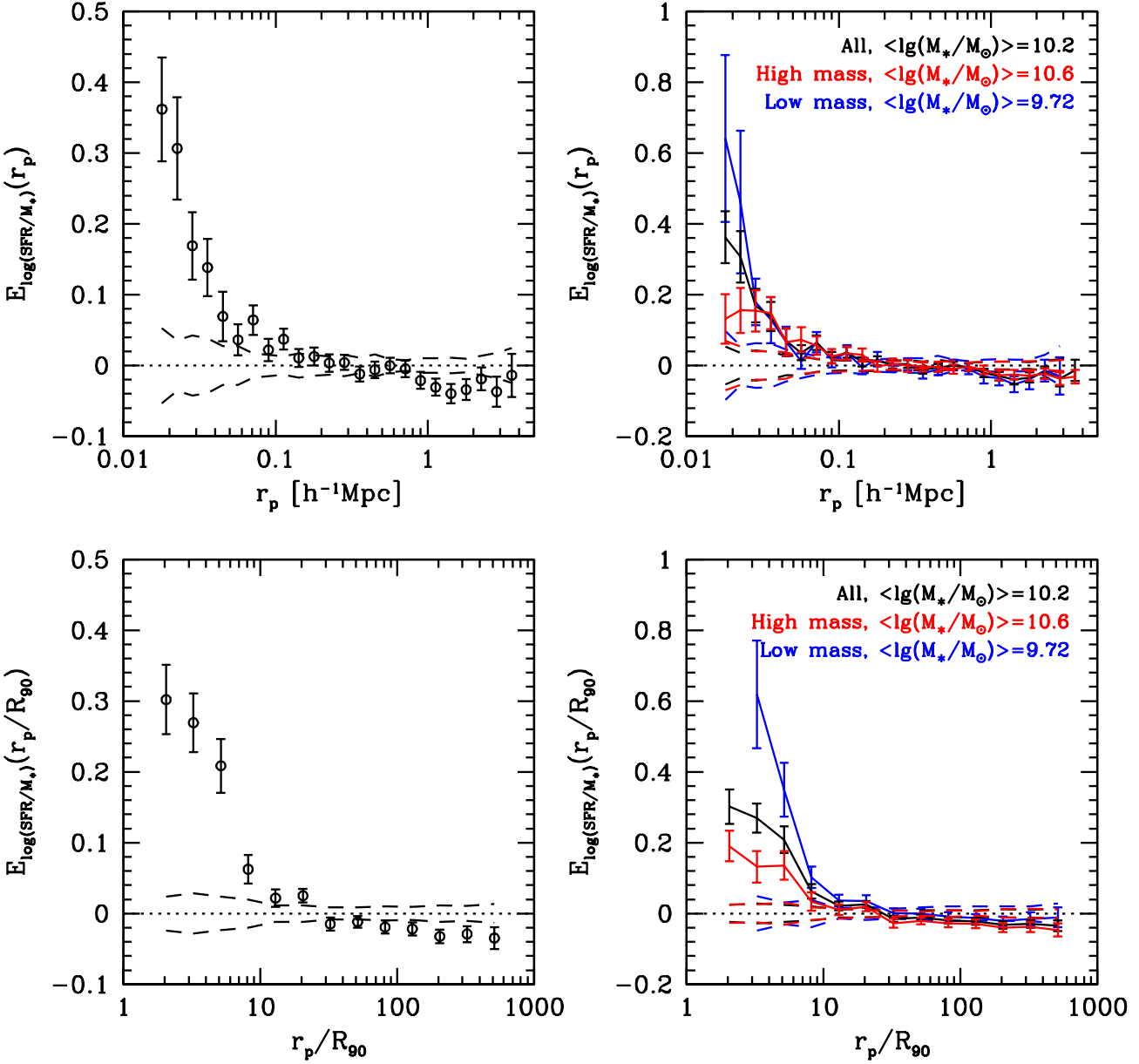,clip=true,width=\textwidth}}
\caption{Star  formation enhancement  as a  function of  the projected
separation  $r_p$  (top  panels)  and  as a  function  of  the  scaled
separation  $r_p/R_{90}$ (bottom panels),  for all  the high  S/N star
forming galaxies  (left panels) and for galaxies  in different stellar
mass ranges  (right panels).  All the errors  are estimated  using the
Bootstrap  resampling  technique.   The  dashed lines  in  each  panel
indicate  the  variance  between  10  realizations in  which  the  sky
positions of  the star-forming galaxies  are randomized. See  the text
for details.}
\label{fig:sfef}
\end{figure*}

\begin{figure*}   
\centerline{\psfig{figure=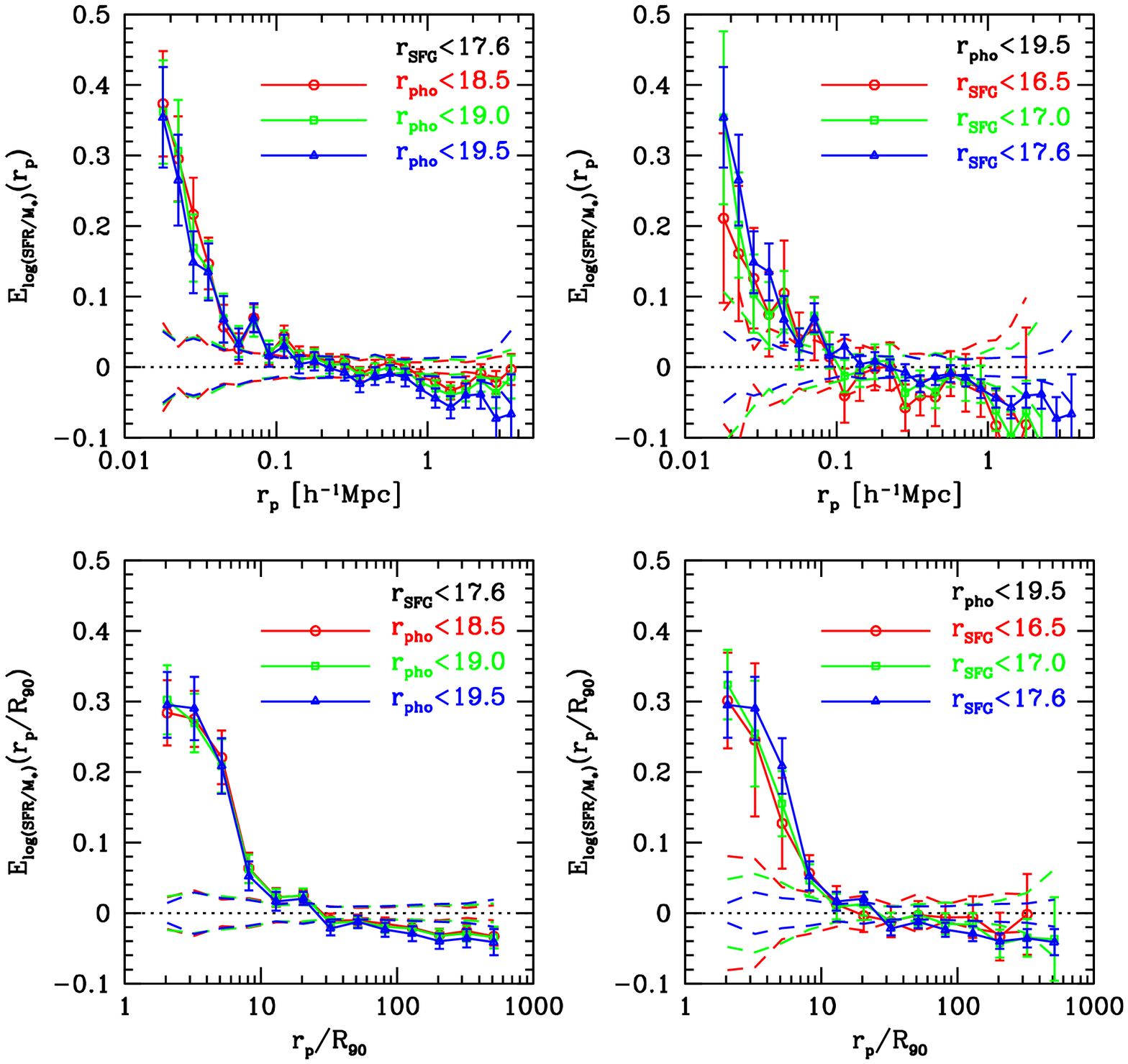,clip=true,width=\textwidth}}
\caption{Star  formation  enhancement   as  a  function  of  projected
separation  $r_p$  (top  panels)  and  as a  function  of  the  scaled
separation  $r_p/R_{90}$  (bottom panels).  In  the left-hand  panels,
different  symbols connected  by solid  lines correspond  to reference
samples with  different limiting magnitudes (as  indicated), while the
magnitude    of   star-forming   samples    is   kept    constant   at
$r_{SFG}=17.6$.  In the  right-hand  panels, the  reference sample  is
always  limited  at $r_{pho}=19.5$  but  the  magnitude  limit of  the
star-forming  sample is changed  (as indicated).   All the  errors are
estimated using the Bootstrap  resampling technique.  The dashed lines
in each panel  indicate the variance between 10  realizations in which
the sky  positions of the  star-forming galaxies are  randomized.  See
the text for details.  }
\label{fig:sfef_fainter}
\end{figure*}

In this section, we probe  the relationship between star formation and
galaxy interactions by quantifying the enhancement in star formation as
a function of the projected  separation between two galaxies.  We also
study how  the enhancement depends  on the physical properties  of the
main galaxy.  We compute how the average value of $SFR/M_\ast$ changes
as a  function of the projected  distance to the  neighbours. From now
on, we restrict our attention to the sample of high $S/N$ star-forming
galaxies. These  lie in low density environments  where processes such
as ram-pressure stripping, gas starvation  etc should play a much less
important role (see Figure 4).

The  neighbours  of a  galaxy  are  identified  using the  photometric
reference sample.   The advantage of  using the photometric  sample is
that the result is not affected by incompleteness (e.g.  the effect of
fibre collisions).  However, the disadvantage is that some fraction of
the  close neighbours  will not  be  true nearby  systems, but  rather
chance  projections of  foreground  and background  galaxies that  lie
along the line-of-sight. We correct  for this as follows: We count the
number  of  companions  in  the  photometric  reference  sample  at  a
projected  physical distance  $r_p$ for  each galaxy  with a  high S/N
measure of  the specific  star formation rate  $\log(SFR/M_\ast)$.  We
also generate  10 random  samples that have  the same geometry  as the
photometric reference  sample by randomizing  the sky position  of the
photometric  objects and keeping  all the  other quantities  (e.g. the
magnitudes) fixed. We use these random catalogues to estimate the mean
number of projected companions  expected at random around each galaxy.
The  true number of  companions at  separation $r_p$  is given  by the
difference  between the  observed and  the projected  random companion
count.  We  then calculate a weighted average  specific star formation
rate at projected distance $r_p$  by weighting each galaxy by its true
companion number.  The  enhancement in $\log(SFR/M_\ast$), $E_X(r_p)$,
is  defined as  the difference  between the  weighted average  and the
unweighted one.  This can be written as
\begin{equation}
E_X(r_p) = \frac{\sum_i^NX_i[n_{o,i}(r_p)-n_{p,i}(r_p)]}
{\sum_i^N[n_{o,i}(r_p)-n_{p,i}(r_p)]}-\frac{\sum_i^NX_i}{N},
\end{equation} where $X_i=\log(SFR_i/M_{\ast,i})$ is the specific star
formation rate of the i'th galaxy, and $n_{o,i}$ and $n_{p,i}$ are the
observed and projected random companion counts as described above.

We  first consider all  high S/N  star-forming galaxies  with $r$-band
apparent magnitude in the  range $14.5<r<17.6$.  To begin, we restrict
the photometric reference sample  to galaxies with $r$-band magnitudes
$r<19.0$. In order to ensure that we are finding similar neighbours at
all  redshifts, we  only  consider neighbouring  galaxies that  are
brighter   than   $r_{SFG}+1.4$   mag.   The  result   is   shown   in
Figure~\ref{fig:sfef}.  The  errors   are  estimated  using  Bootstrap
resampling techniques. The dashed  lines indicate the variance between
10 samples in which we randomize the sky positions of the star-forming
galaxies.

\begin{figure*}
\centerline{\psfig{figure=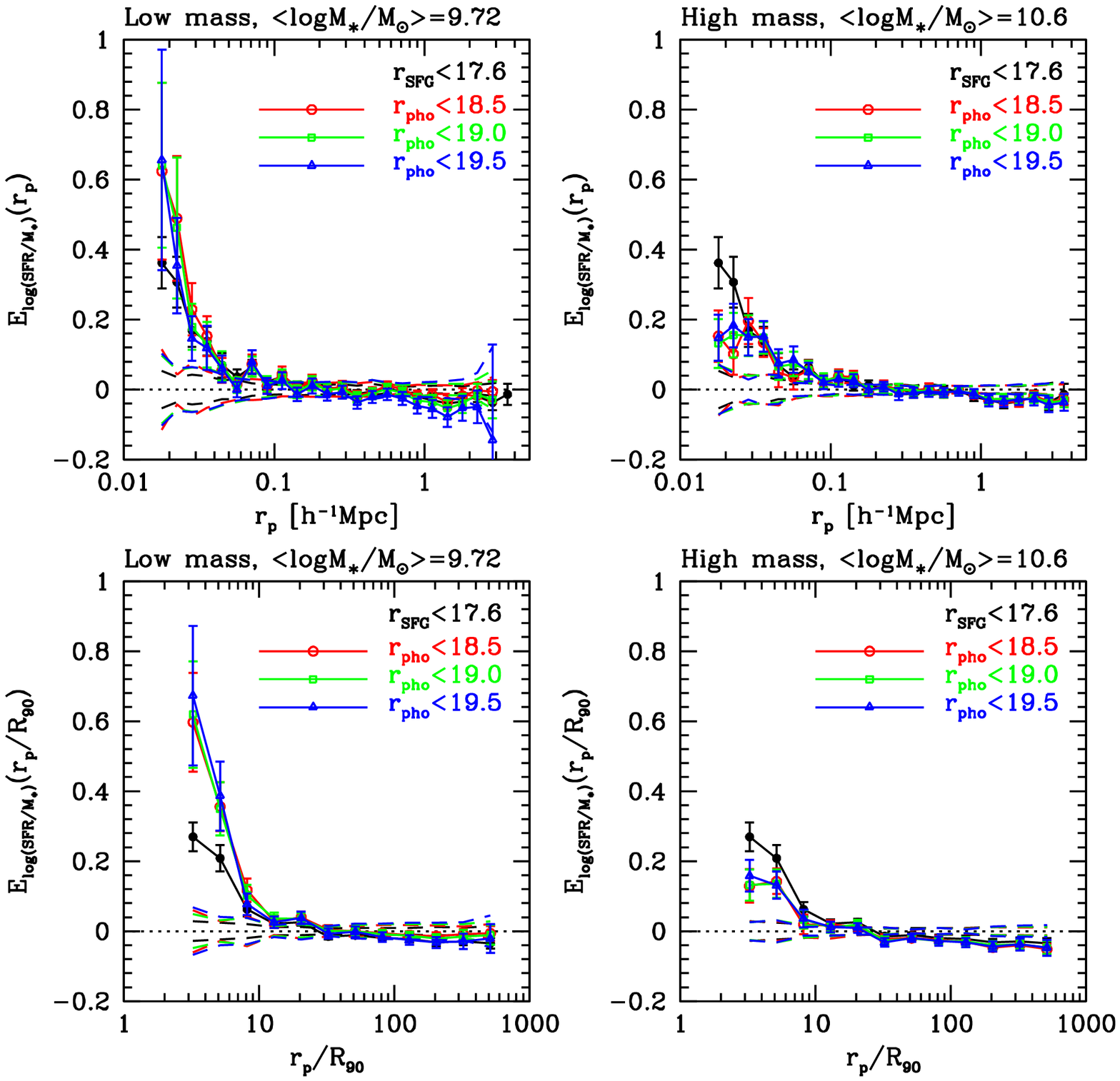,clip=true,width=\textwidth}}
\caption{The     same     as      the     bottom-left     panel     of
Figure~\ref{fig:sfef_fainter},  but for  the low  mass  (the left-hand
panel) and the high mass (the right-hand panel) subsamples separately.
To  guide  the  eye, the  result  for  the  whole  sample in  case  of
$r_{SFG}<17.6$ and $r_{pho}<19.0$ is plotted as solid black circles in
every panel.  }
\label{fig:sfef_smass_fainter}
\end{figure*}

\begin{figure*}
\centerline{\psfig{figure=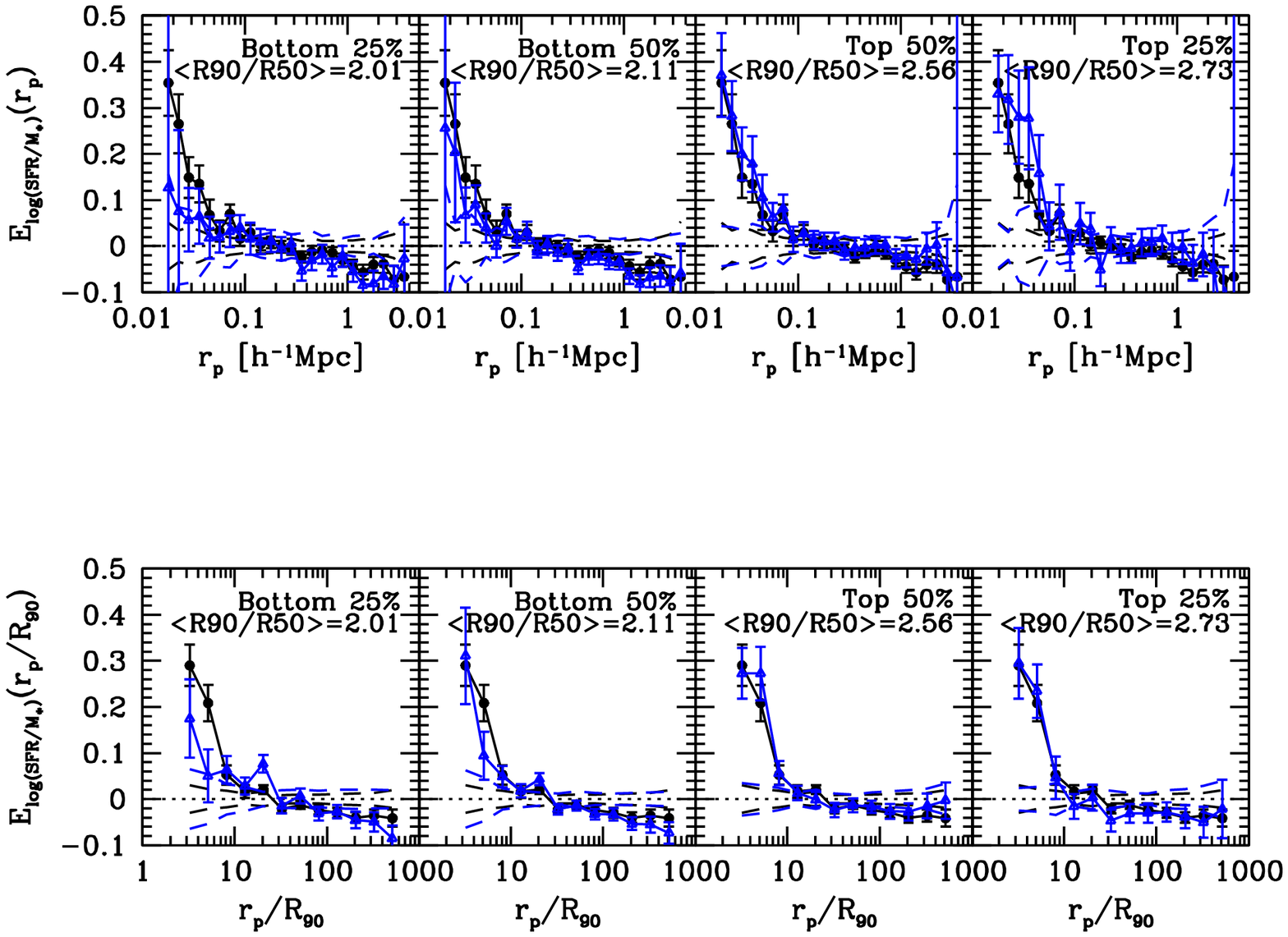,clip=true,width=\textwidth}}
\caption{Similar to  the previous  plot but for  star-forming galaxies
with different  concentration indices, as indicated  above each panel,
and for the  case of $r_{SFG}<17.6$ and $r_{pho}<19.5$  only. The blue
triangles  are for the  concentration subsamples  and the  black solid
circles are for the whole sample.}
\label{fig:sfef_c}
\end{figure*}

The top left  panel of Figure~\ref{fig:sfef} gives the  result for the
sample as a whole. On scales larger than a Mpc or so $E_X$ is constant
at a  slightly but significantly  negative value. This is  because the
average in equation 1 is  pair-weighted, and galaxies in massive halos
have  lower  specific star  formation  rates  than  average, but  more
"companions" at  large $r_p$ than  average as a result  of large-scale
bias effects. On  scales below about 100 kpc,  $E_X$ increases sharply
and reaches values corresponding to a factor of about two at projected
separations  less   than  20   kpc.   In  the   top  right   panel  of
Figure~\ref{fig:sfef}, we  plot results for galaxies  divided into two
different ranges in stellar mass.   These results show that there is a
strong  dependence of star  formation enhancement  on galaxy  mass, in
that star formation  in small galaxies is more  strongly enhanced at a
given projected separation.

In the bottom  panels, we scale the projected  separation $r_p$ by the
physical size of  the galaxy.  We use $R_{90}$,  the radius containing
90\%  of  the  total  $r$-band  light, to  calculate  a  {\em  scaled}
projected  separation  and recompute  the  enhancement  function as  a
function of  this scaled quantity.  One  can see that  the results are
quite similar.   Star formation is  enhanced at separations  less than
$\sim 10$  times the optical  radius of the  galaxy and the  effect is
stronger for lower mass systems.

We  now  investigate  the  importance  of the  relative  mass  of  the
companion galaxy in determining the  degree to which star formation is
enhanced  in  the primary  galaxy  by  analyzing  galaxy samples  with
different limiting  magnitudes.  We first keep the  magnitude limit of
star-forming  sample  constant  at  $r_{SFG}=17.6$, and  explore  what
happens if we  change the limiting magnitude of  the reference sample.
The results are plotted as circles for $r_{pho}<18.5$ and as triangles
for     $r_{pho}<19.5$     in      the     left-hand     panels     of
Figure~\ref{fig:sfef_fainter}.   The  result  shown  in  the  previous
figure is plotted as squares.   Next, we fix the limiting magnitude of
the  reference sample  at $r_{pho}=19.5$,  but decrease  the magnitude
limit of  the star-forming sample.  The maximum allowed  difference in
magnitude between  the star-forming galaxy  and its companion  is also
increased accordingly.  Results are shown for $r_{SFG}<16.5, 17.0$ and
$17.6$.  in the right-hand panels of Figure~\ref{fig:sfef_fainter}.

\begin{figure}
\centerline{
\psfig{figure=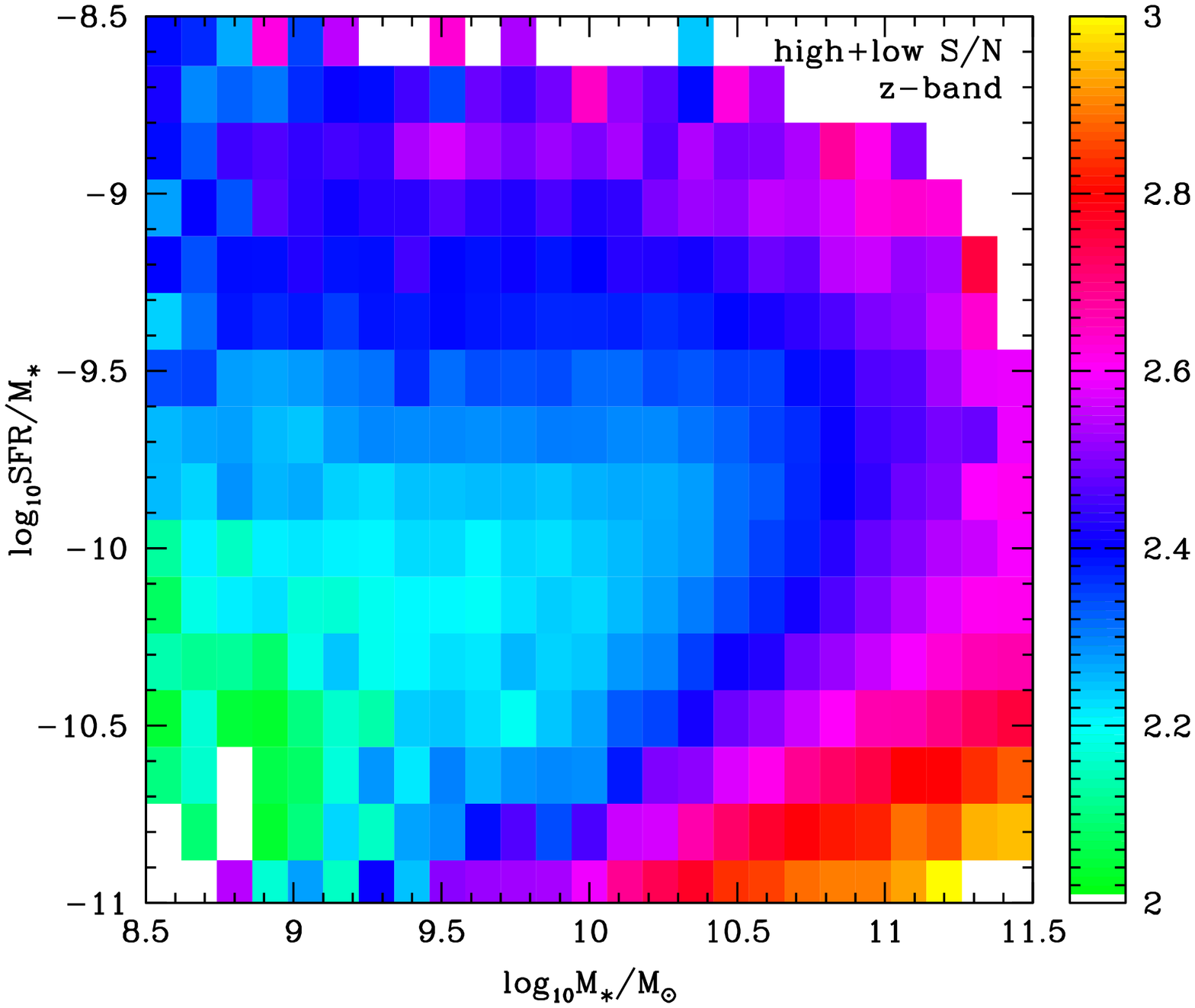,clip=true,width=0.5\textwidth}}
\caption{Distribution of both  high and low S/N star-forming galaxies
in  the plane  of stellar  mass versus  specific star  formation rate,
coloured by  concentration index  $R90/R50$ measured in  the $z$-band.
The color coding  of $R90/R50$ is shown in the  bar at the right-hand.}
\label{fig:con_contour}
\end{figure}

We  see  from Figure~\ref{fig:sfef_fainter}  that  the star  formation
enhancement  depends  very  little  on  the  mass  ratio  between  the
star-forming galaxy and its companion.  There are small changes in the
expected  direction  (i.e.  there  is  slightly  less enhancement  for
companions  with  lower  relative   mass),  but  to  first  order  the
enhancement  function remains remarkably  constant for  different mass
ratios.    In  Figure~\ref{fig:sfef_smass_fainter},   we   divide  the
star-forming  sample into  two  different stellar  mass intervals  and
explore if our results change.   We find that the enhancement function
has very little dependence on the mass ratio of the companion for both
low mass and high mass star-forming galaxies.

Finally,  we investigate  the enhancement  function for  galaxies with
different  structural   properties.   We  divide  all   the  high  S/N
star-forming  galaxies  into   different  intervals  of  concentration
parameter  $C$ and  repeat the  analysis described  above for  each of
these subsamples.  In Figure~\ref{fig:sfef_c}, we plot the results for
star-forming galaxies  with $r_{SFG}<17.6$ and  for reference galaxies
with $r_{pho}<19.5$.  We  see that the  star formation
enhancement does depend  on $C$, in that the  galaxies with larger $C$
values are  more strongly enhanced. One possible  explanation for this
effect  is that  interaction-induced starbursts  occur when  gas flows
into  the  core of  a  galaxy, causing  it  to  become more  centrally
concentrated \citep{Sanders-Mirabel-96}.

In Figure~\ref{fig:con_contour}, we  investigate how the concentration
index of a star-forming galaxy depends on its location in the plane of
specific  star formation  rate versus  stellar mass.   We see  that at
fixed  stellar mass, the  average concentration  index is  highest for
galaxies that are  currently experiencing both higher-than-average and
lower-than-average  rates of  star formation.   One  interpretation of
this plot is  that tidal interactions cause gas to  flow from the disk
to the  nucleus and  this triggers  a starburst at  the centre  of the
galaxy. The formation  of new stars in the  central regions causes the
concentration index to  increase. The starburst is then  followed by a
period of relative quiescence, which lasts until the galaxy is able to
accrete more gas into its disk.  Formation of stars in the disk brings
the galaxy  back into  the "central plane"  occupied by  galaxies with
$\log  SFR/M_*  \sim -9.5$  in  Figure~\ref{fig:con_contour}.  In  the
bottom-right corner  of the plot, both stellar  mass and concentration
are high,  but the specific star  formation rate is low.   This is the
regime of early-type galaxies.

\section{Close neighbour counts}

\begin{figure*}
\centerline{
\psfig{figure=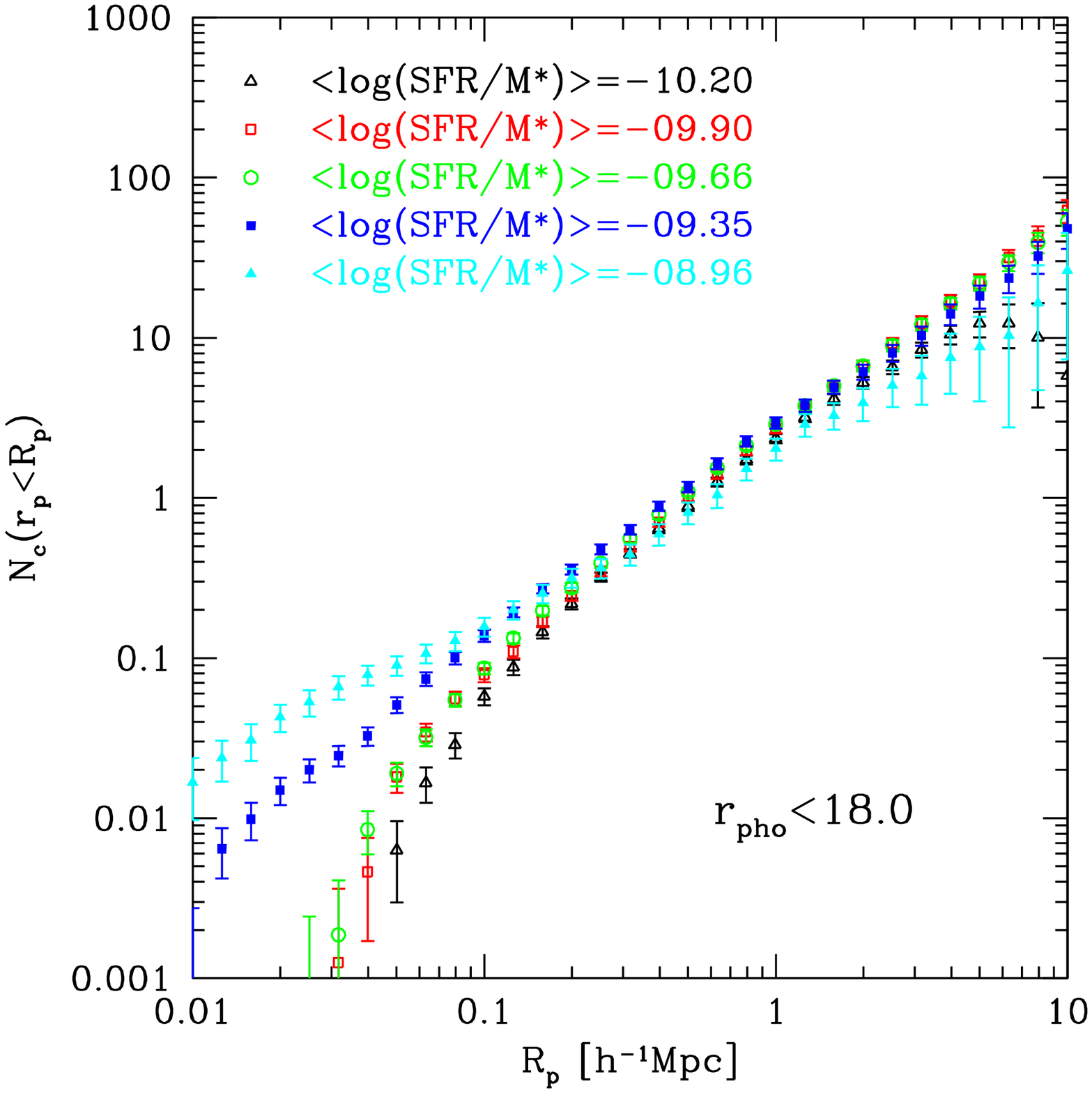,clip=true,width=0.33\textwidth}
\psfig{figure=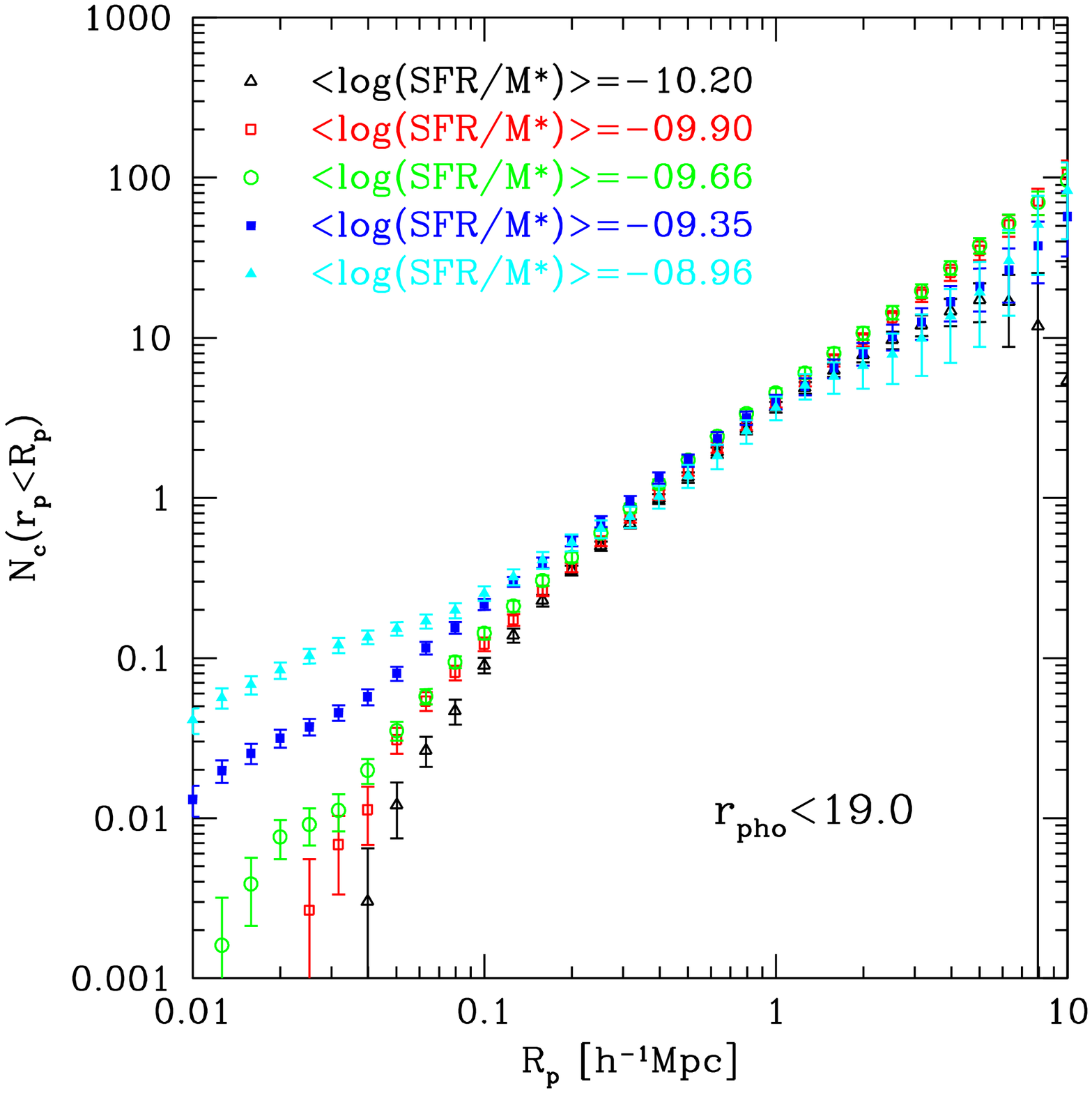,clip=true,width=0.33\textwidth}
\psfig{figure=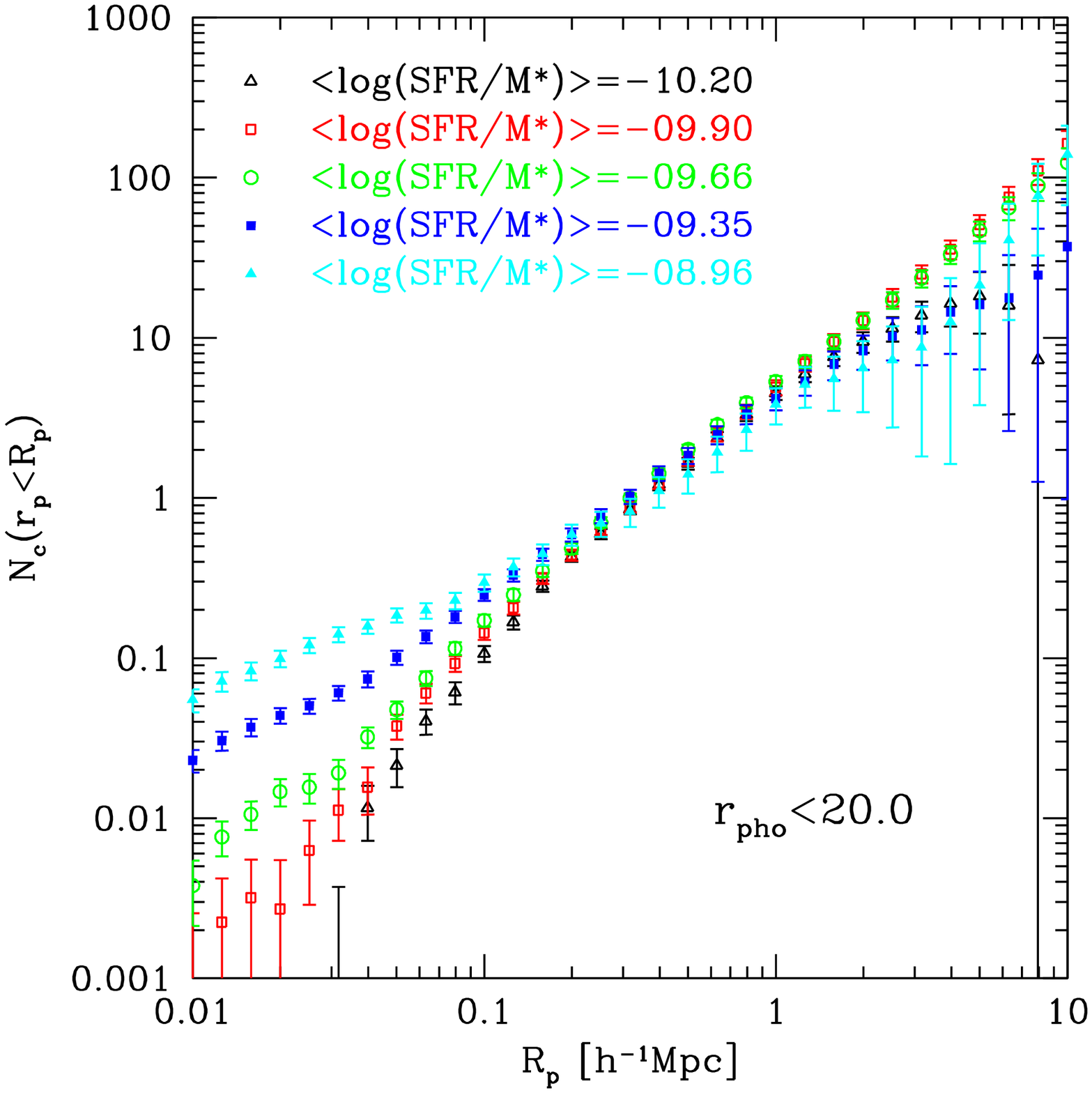,clip=true,width=0.33\textwidth} }
\caption{Average counts of galaxies  in the photometric sample (panels
from left  to right: $r_{lim}  < 18$, $19$,  and $20$) within  a given
projected  radius  $R_p$ from  the  star-forming galaxies.   Different
symbols  are  for  star-forming  galaxies in  different  intervals  of
specific star formation rate, as indicated.}
\label{fig:counts}
\end{figure*}

In  this section, we  investigate whether  tidal interactions  are not
only  a sufficient, but  also a  necessary condition  for a  galaxy to
experience enhanced  star formation.  We count the  number of galaxies
in the photometric sample in  the vicinity of the star-forming galaxies
and make a statistical correction for the effect of chance projections
by subtracting the average count around randomly placed galaxies.

In  Figure~\ref{fig:counts} we plot  the average  correlated neighbour
count  (i.e. after statistical  correction for  uncorrelated projected
neighbours)   within   a  given   value   of   the  projected   radius
$R_p$.  Results  are  shown  for  high S/N  star-forming  galaxies  in
different intervals  of specific star formation rate.  We have trimmed
each  subsample  so that  they  each  have  the same  distribution  in
redshift  and  in stellar  mass  $M_*$.   Panels  from left  to  right
correspond  to  photometric  reference  samples that  are  limited  at
$r=18.0,  19.0$ and  $20.0$.   The star-forming  sample  always has  a
limiting magnitude of 17.6.

\begin{figure}
\centerline{\psfig{figure=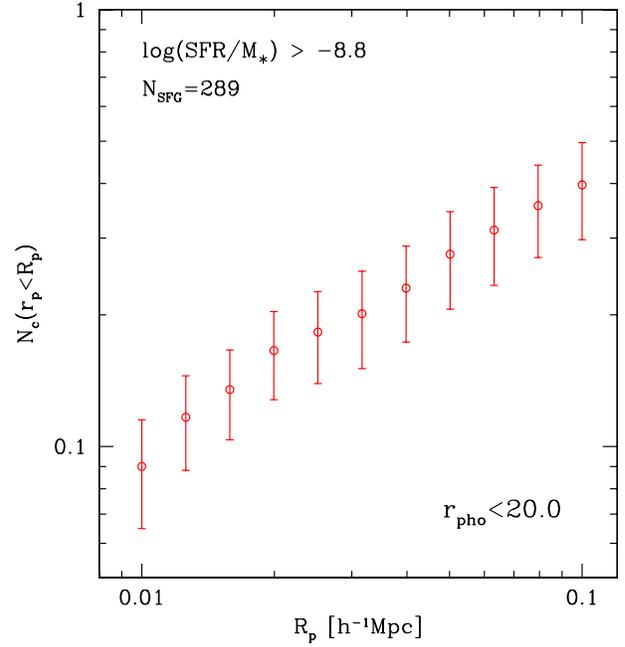,clip=true,width=0.5\textwidth}}
\caption{Same as the  right-hand panel of Figure~\ref{fig:counts}, but
for 289 galaxies  that have the highest specific  star formation rates
($\log_{10}(SFR/M_\ast)>-8.8$).   Results are  shown  only for  scales
below 100kpc.}
\label{fig:counts_highest_bin}
\end{figure}

\begin{figure*}
\centerline{\psfig{figure=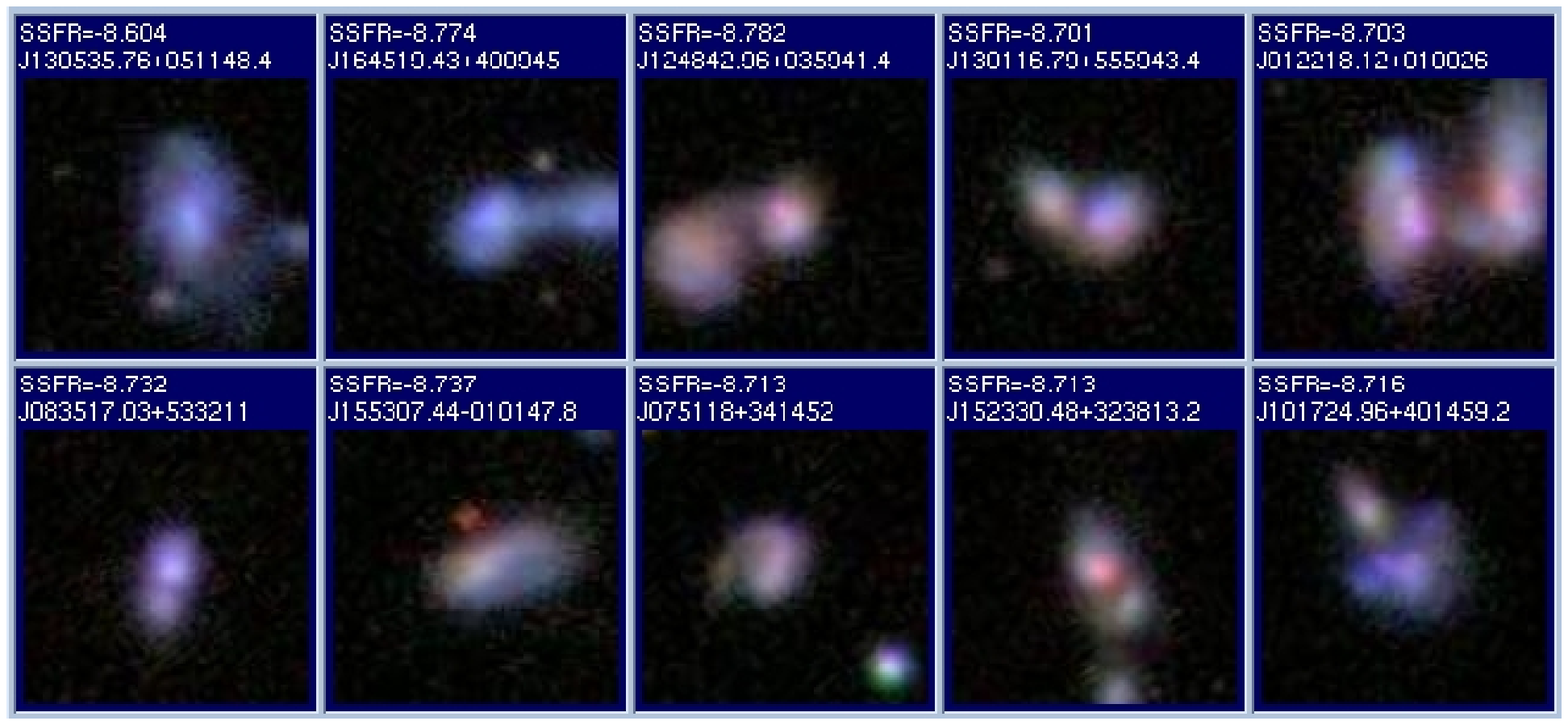,clip=true,width=\textwidth}}
\centerline{\psfig{figure=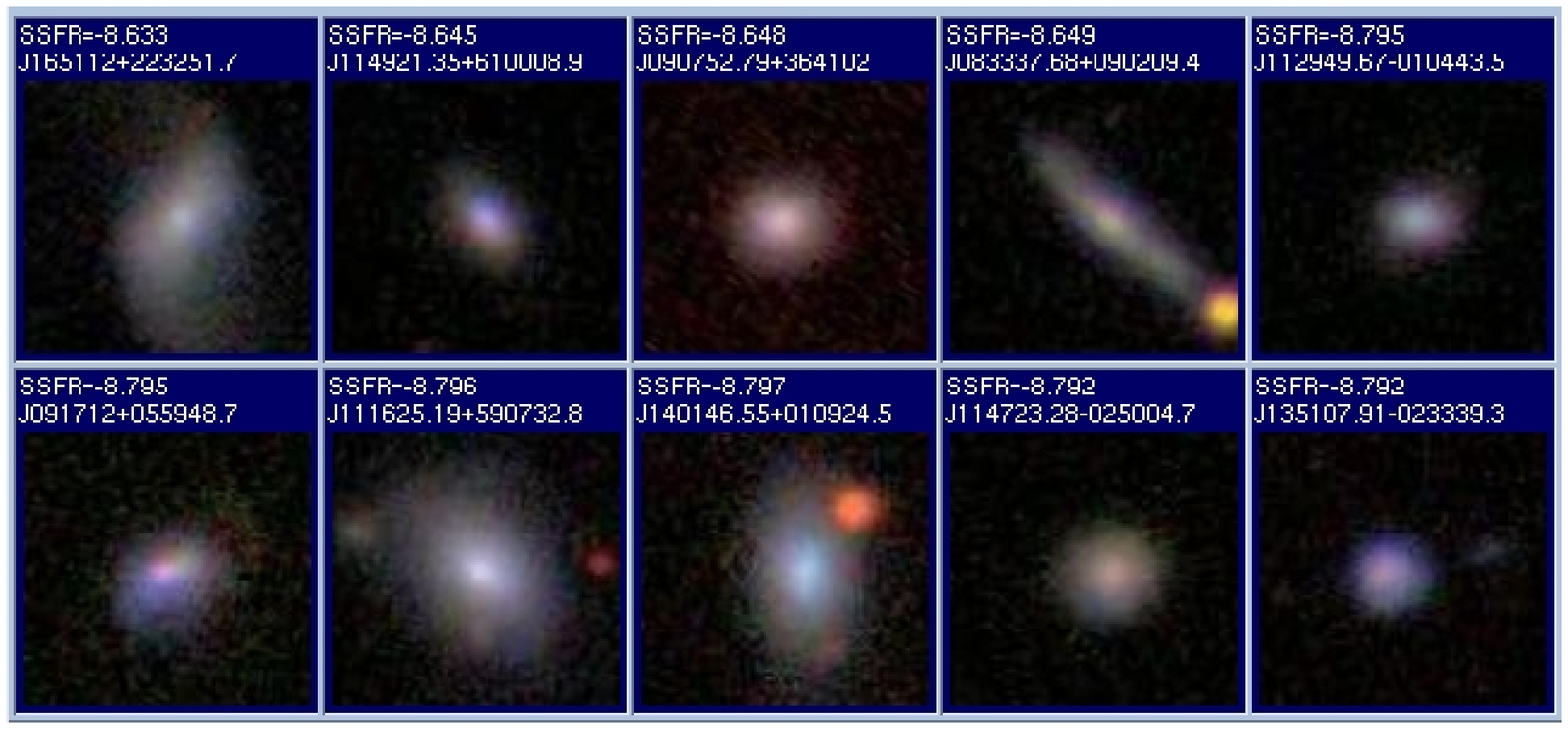,clip=true,width=\textwidth}}
\caption{SDSS  optical image  for  20 star-forming  galaxies that  are
included  in   the  subsample  of  highest   specific  star  formation
rate($\log_{10}(SFR/M_\ast)>-8.8$).  Images are  shown for 10 galaxies
that are classified as "mergers" (top panels) and for 10 galaxies that
are classified as "no mergers" (bottom panels).}
\label{fig:images}
\end{figure*}

Figure~\ref{fig:counts} shows that  the counts around the star-forming
galaxies with  different specific star  formation rates match  well on
large scales.  On scales smaller than $\sim 100$ kpc, there are strong
trends  in  the number  of  neighbours  as  a function  of  SFR/$M_*$;
galaxies with  higher star formation rates  are more likely  to have a
near neighbour.

It is interesting  that the average number of  close neighbours around
galaxies with  low-to-average values of $SFR/M_\ast$ is  close to zero
on scales  less than  20-30 kpc.   On scales less  than 100  kpc, only
around  3\%  of  the galaxies  in  the  lowest  SFR/$M_*$ bin  have  a
companion.  This implies that tidal interactions that do not result in
enhanced  star formation are  a rare  occurrence.  This  is consistent
with the findings  of \citet{DiMatteo-07} who find that  about 85\% of
their simulated sample of  interacting galaxies show an enhancement in
star formation by a factor $>$ 2.  As the specific star formation rate
increases, the average number of close neighbours also rises.  Fifteen
percent  of galaxies  in our  highest SFR/$M_*$  bin have  a companion
within  20-30 kpc  and  this rises  to a  value  close to  30\% if  we
consider  companions  within 100  kpc  from  the  primary galaxy.   In
Figure~\ref{fig:counts_highest_bin},  we  show   the  result  for  289
galaxies with  the very highest  specific star formation rates  in our
sample.  The fraction of galaxies that have a companion within 100 kpc
increases to values close to 40\%.  However, it is still true that not
{\em every} star-bursting galaxy in our sample has a close companion.

How can we  explain those galaxies with very  high $SFR/M_\ast$ but no
close neighbours?  We have  visually examined the SDSS $r$-band images
of  160  star-forming  galaxies  that  are  included  in  the  highest
$SFR/M_\ast$  subsample but  have  no companions  within  50 kpc.   We
classified the  systems according  to whether or  not they  show clear
signs of  mergers or interactions,  including double nuclei  and tidal
tails.   Three  of   us  (CL,  GK  and  Roderik   Overzier)  did  this
independently to make sure that  we obtained the same answer.  We find
that more  than half  of such galaxies  show clear evidence  of recent
mergers.   In Figure~\ref{fig:images},  we show  some examples  of our
classifications.  We  thus conclude that,  at least for  galaxies with
very  high  $SFR/M_\ast$, interactions  or  mergers  are the  dominant
mechanism for triggering and enhancing their star formation.

\section{Effect of rich environments}

\begin{figure*}
\centerline{\psfig{figure=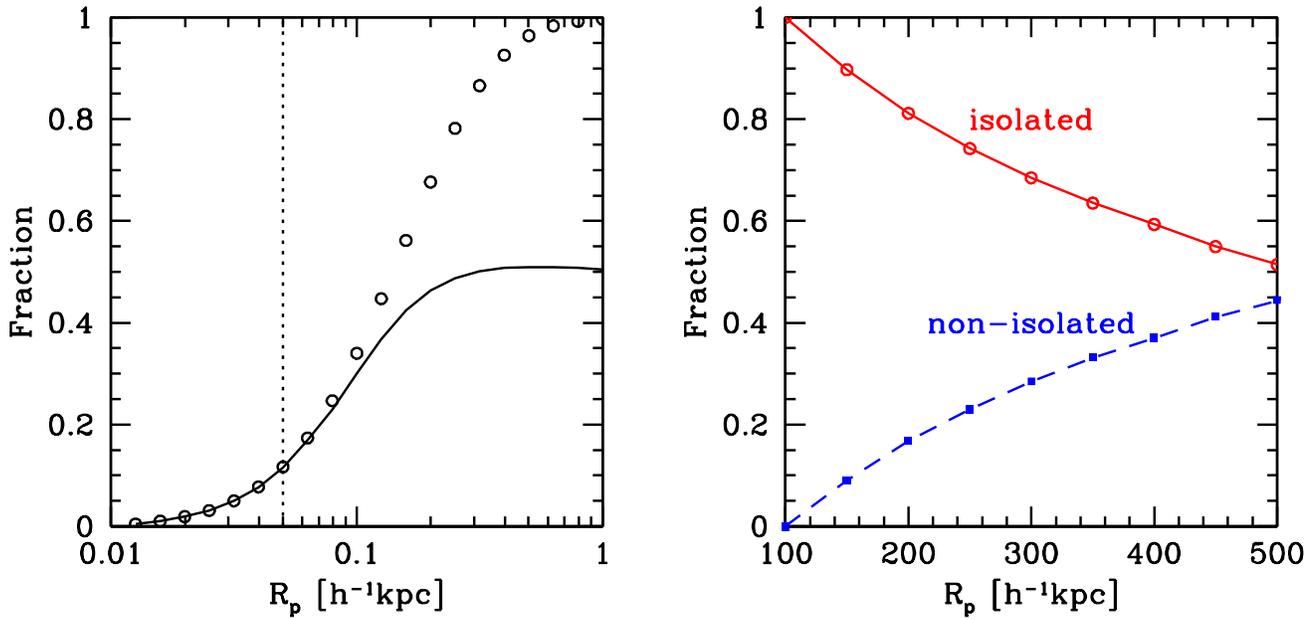,clip=true,width=\textwidth}}
\caption{Left: Ope circles show  the fraction of star-forming galaxies
that have  at least one companion  in the photometric  sample within a
given projected  radius $R_p$. The  solid line shows the  result after
the number of companions is corrected using random samples. Right: the
fraction  of   "isolated"  (red)  and   "non-isolated"  (blue)  paired
galaxies, classified according to  whether they have companions in the
spectroscopic reference sample with  the projected separation $r_p$ in
the range  100 $h^{-1}$kpc  $< r_p <  $ $R_p$ and  velocity difference
smaller than 500 km s$^{-1}$.  The parent sample of paired galaxies is
selected from all the high  S/N star-forming galaxies by requring that
a paired galaxy  has at least one companion  within a projected radius
of  50 $h^{-1}$kpc  in  the photometric  reference  sample limited  at
$r_{pho}=19.0$. Solid/dashed lines (open circles/squares) show results
obtained with the corrected (observed) companion numbers.}
\label{fig:pfraction}
\end{figure*}

\begin{figure*}
\centerline{\psfig{figure=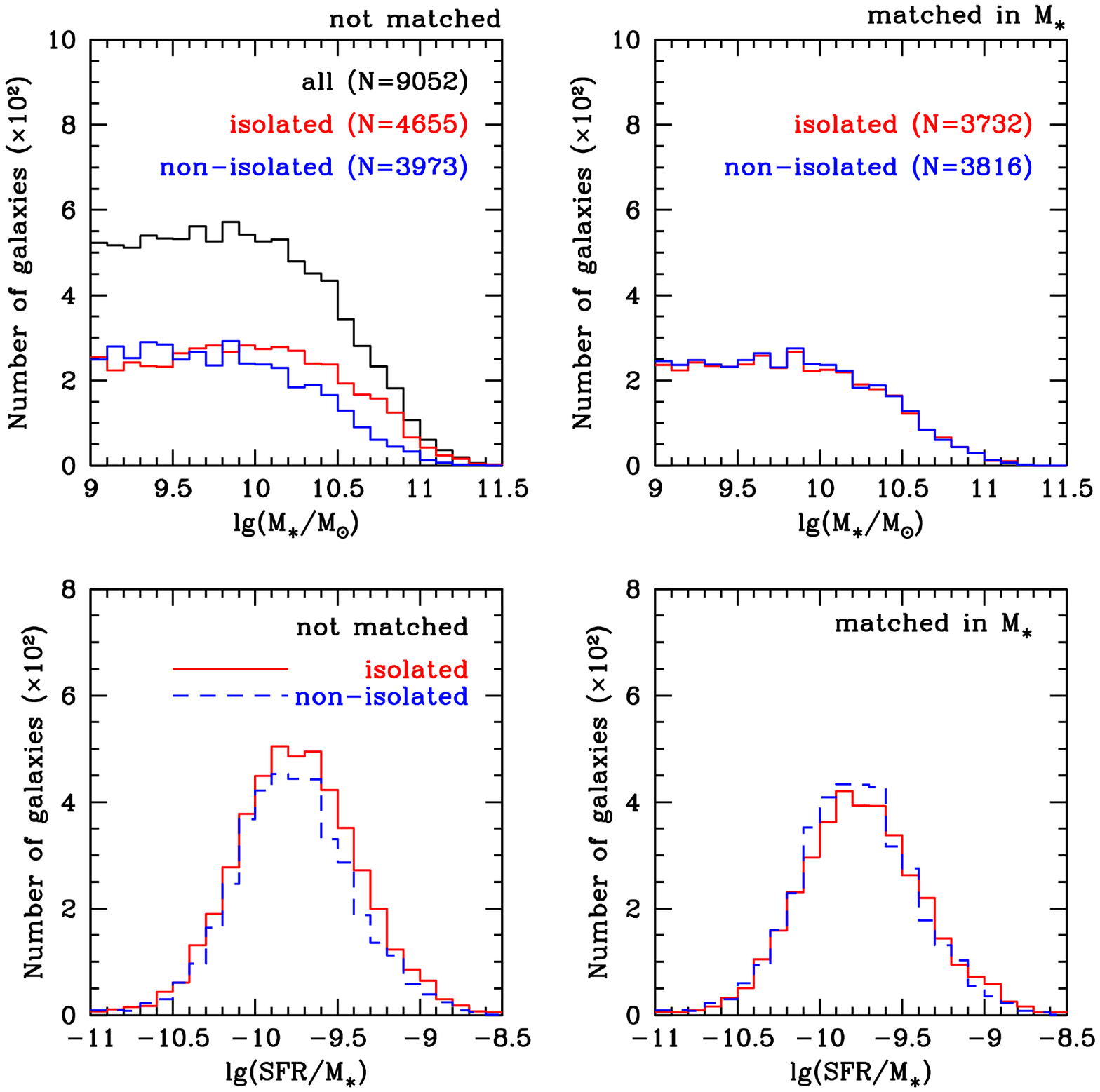,clip=true,width=\textwidth}}
\caption{Distribution of specific star formation rate for the "isolated"
(solid) and "non-isolated" (dashed) paired galaxies. In the right-hand panel,
the two samples are matched closly in stellar mass.}
\label{fig:sfr_hist}
\end{figure*}

Fig.~\ref{fig:sfef}   (\S~\ref{sec:enhancement})    shows   that   the
enhancement  in  $\log(SFR/M_\ast)$  on  large  scales  is  not  zero.
Rather, it  is constant at  a small but significantly  negative value.
We   have  attributed   this  large-scale   bias  to   the  well-known
anti-correlation  of the  star formation  rates in  galaxies  with the
richness of their local  environment. The question then arises whether
the  star formation enhancements  that we  compute may  not be  a true
reflection  of the effect  of galaxy-galaxy  interactions, but  may be
biased because some close pairs  are not real interacting systems, but
are associated with group/cluster environments.

In order  to address this problem,  we have selected a  sample of 9052
paired galaxies from all the  high S/N star-forming galaxies. A galaxy
is defined  to be  paired if it  has at  least one companion  within a
projected radius of 50 $h^{-1}$kpc in the photometric reference sample
limited at  $r_{pho}=19.0$. We use  the the photometric  sample rather
than the  sepectroscopic one  to select pairs,  so that  the resulting
sample is  not biased to  pairs with approximately equal  masses.  The
maximum  pair separation  is chosen  to yield  a sample  in  which the
contamination   by  chance   projections  is   negligible.    This  is
demonstrated in the  left-hand panel of Fig.~\ref{fig:pfraction}.  The
fraction of star-forming galaxies that  have at least one companion in
the  photometric  sample within  a  given  projected  radius $R_p$  is
plotted as a function of $R_p$ as open circles, and is compared to the
result (the  solid line) after  the number of companions  is corrected
with  the help  of  random samples  (see \S~\ref{sec:enhancement}  for
details). As can be seen,  at separations smaller than 50 $h^{-1}$kpc,
the correction is negligible.

Next,  We  classify the  paired  galaxies  as  either ``isolated''  or
``non-isolated''  according to  whether  they have  companions in  the
spectroscopic reference sample with  projected separation $r_p$ in the
range 100 $h^{-1}$kpc $< r_p < $ $R_p$ and velocity difference smaller
than 500 km s$^{-1}$. A galaxy is isolated if this ring-like region is
completely empty.  In contrast, a  non-isolated galaxy is  required to
have at least one companion in  the annulus that is brighter than that
galaxy    in     the    $r$-band.    The     right-hand    panel    of
Fig.~\ref{fig:pfraction} shows how the  fraction of these two types of
paired galaxies changes as  the maximum separation $R_p$ increases. We
see that  the sample is  always dominated by isolated  pairs. However,
even a  small fraction of  non-isolated galaxies may still  change the
enhancement function significantly,  because the quantities defined in
equation 1  are pair-weighted and galaxies in  richer environment have
more companions. In the  left-hand panel of Fig.~\ref{fig:sfr_hist} we
compare  the distribution  of specific  star formation  rates  for the
isolated and the non-isolated  samples. The right-hand panel shows the
results after  the two samples are  matched in stellar  mass with each
other (This  is important because massive galaxies  are more clustered
than  less massive  galaxies  \citep{Li-06a}).  As  can  be seen,  the
effect of  a rich  environment is quite  small.  The  average specific
star formation  rate of the isolated  sample differs from  that of the
non-isolated sample by  $\la$ 0.03 dex only.  This  is consistent with
the findings of \citet{Balogh-04} who showed that the main effect of a
dense  environment on  large  scales  is that  the  {\em fraction}  of
emission-line  galaxies  decreases,   but  that  the  distribution  of
equivalent  widths  of  H$\alpha$  among the  star-forming  population
remains virtually  constant.  Here we  show that the same  thing holds
for galaxies with  close pairs. We thus conclude  that the large-scale
environment does  not affect the measured  star formation enhancements
in our  sample of high S/N  star-forming galaxies on  scales below 100
$h^{-1}$kpc.

\section{Summary and Discussion}

We  find  that  the  clustering  amplitude of  high  S/N  star-forming
galaxies  depends strongly  on  the specific  star  formation rate  on
scales less than 100  kpc. The clustering amplitude increases smoothly
as a function of SFR/$M_*$ and the increase in amplitude is largest at
the smallest  projected separations.   We interpret this  behaviour as
the signature of  tidal interactions, which lead to  inflow of gas and
an enhancement in star formation  in the two interacting galaxies.  At
low  values of  SFR/$M_*$, the  clustering amplitude  again increases.
The increase occurs on {\em all scales} and probably reflects the fact
that star formation  in galaxies switches off after  they are accreted
onto larger structures such as groups and clusters.

We have explored how the  average star formation rates of galaxies are
enhanced  as   a  function  of  the  projected   separation  of  their
companions.   The  enhancement is  a  strong  function of  separation,
increasing  from zero  at  $r_p >  100  kpc$ to  factors  of 1.5-4  at
$r_p=20$ kpc.   We find  that the enhancement  at given  separation is
stronger  for  lower mass  galaxies.   Remarkably,  we  find that  the
enhancement has almost no dependence on the relative luminosity of the
companions.

The tidal force  between two objects is expected  to scale as $d^{-3}$
and $m/M$,  where $d$ is  the separation and  $m/M$ is the  mass ratio
between the two  objects, so it is perhaps not  surprising that we see
stronger star  formation enhancement as a function  of separation than
as a  function of  mass ratio.  Nevertheless,  in this study,  we find
that the effect  of a companion that is 3  magnitudes fainter than the
primary galaxy is  very similar to that of a companion  that is only a
factor of 3  less luminous than the primary.   This is quite startling
and is worthy of further investigation.

In  order   to  explore  whether  tidal  interactions   are  not  only
sufficient, but also a necessary condition for enhanced star formation
in a  galaxy, we have computed background  subtracted neighbour counts
around the galaxies in our sample.  We find that the average number of
galaxies around galaxies  with low values of SFR/$M_*$  is very small.
At the very  highest specific star formation rates,  more than 40\% of
the galaxies in our sample  have a companion within a projected radius
of 100 kpc.  Visual inspection of the high  SFR/$M_*$ galaxies without
companions reveals that more than  50\% of these are clear interacting
or merging systems.  We thus  conclude that tidal interactions are the
primary mechanism for inducing the  highest rates of star formation in
galaxies in the local Universe.

Finally, we find clear evidence  that tidal interactions not only lead
to  enhanced star  formation,  but also  cause  structural changes  in
galaxies.  Many of  the  most strongly  star-forming  galaxies in  our
sample  have   concentration  indices  similar  to   those  of  normal
early-type galaxies.  We note that the concentration index is measured
for the $r$-band  light and not for the stellar  mass, so more careful
analysis  is  needed  before  one  can definitely  conclude  that  the
interactions  will result in  the formation  of a  galaxy with  a high
bulge mass fraction. Nevertheless, we conclude that our results are in
general  accord  with  the  theoretical  picture  first  laid  out  by
\citet{Toomre-Toomre-72},  which showed  that galaxy  interactions can
lead to the growth of bulges and spheroids in galaxies.

\section*{Acknowledgements} 

CL is  supported by the Joint Postdoctoral  Programme in Astrophysical
Cosmology  of  Max  Planck  Institute for  Astrophysics  and  Shanghai
Astronomical Observatory.  CL and YPJ are supported by NSFC (10533030,
10643005, 10633020),  by the Knowledge Innovation Program  of CAS (No.
KJCX2-YW-T05), and by 973  Program (No.2007CB815402).  We are grateful
to the referee for his/her comments which have helped to improve the paper,
and Roderik Overzier for  visually examining and classifying the images
of our  galaxies.  CL, GK and  SW would like to  thank the hospitality
and stimulating atmosphere of the  Aspen Center for Physics while this
work was being completed.

Funding for  the SDSS and SDSS-II  has been provided by  the Alfred P.
Sloan Foundation, the Participating Institutions, the National Science
Foundation, the  U.S.  Department of Energy,  the National Aeronautics
and Space Administration, the  Japanese Monbukagakusho, the Max Planck
Society, and  the Higher Education  Funding Council for  England.  The
SDSS Web  Site is  http://www.sdss.org/.  The SDSS  is managed  by the
Astrophysical    Research    Consortium    for    the    Participating
Institutions. The  Participating Institutions are  the American Museum
of  Natural History,  Astrophysical Institute  Potsdam,  University of
Basel,   Cambridge  University,   Case  Western   Reserve  University,
University of Chicago, Drexel  University, Fermilab, the Institute for
Advanced   Study,  the  Japan   Participation  Group,   Johns  Hopkins
University, the  Joint Institute  for Nuclear Astrophysics,  the Kavli
Institute  for   Particle  Astrophysics  and   Cosmology,  the  Korean
Scientist Group, the Chinese  Academy of Sciences (LAMOST), Los Alamos
National  Laboratory, the  Max-Planck-Institute for  Astronomy (MPIA),
the  Max-Planck-Institute  for Astrophysics  (MPA),  New Mexico  State
University,   Ohio  State   University,   University  of   Pittsburgh,
University  of  Portsmouth, Princeton  University,  the United  States
Naval Observatory, and the University of Washington.


\bsp
\label{lastpage}

\end{document}